\documentclass[aps,prc,preprint,floatfix]{revtex4}

\usepackage{amsmath}
\usepackage{graphicx}
\usepackage{bm}
\DeclareGraphicsExtensions{.ps}

\def\gtorder{\mathrel{\raise.3ex\hbox{$>$}\mkern-14mu
	\lower0.6ex\hbox{$\sim$}}}
\def\ltorder{\mathrel{\raise.3ex\hbox{$<$}\mkern-14mu
	\lower0.6ex\hbox{$\sim$}}}
\newcommand{\eps}{\epsilon}
\newcommand{\veps}{\varepsilon}
\newcommand{\w}{\omega_k}

\preprint{JLAB-THY-09-1011}

\begin{document}

\title{Equivalence of Pion Loops in Equal-Time	\\
	and Light-Front	Dynamics}

\author{Chueng-Ryong Ji}
\affiliation{Department of Physics, Box 8202,
	North Carolina State University,
	Raleigh, North Carolina 27692-8202}
\author{W. Melnitchouk}
\affiliation{Jefferson Lab, 12000 Jefferson Avenue,
        Newport News, Virginia 23606}
\author{A. W. Thomas}
\affiliation{\mbox{Jefferson Lab, 12000 Jefferson Avenue,
        Newport News, Virginia 23606}, and\\
	College of William and Mary, Williamsburg, Virginia 23187}


\begin{abstract}
We demonstrate the equivalence of the light-front and equal-time
formulations of pionic corrections to nucleon properties.
As a specific example, we consider the self-energy $\Sigma$ of a
nucleon dressed by pion loops, for both pseudovector and pseudoscalar
$\pi NN$ couplings.
We derive the leading and next-to-leading nonanalytic behavior of
$\Sigma$ on the light-front, and show explicitly their equivalence in
the rest frame and infinite momentum frame in equal-time quantization,
as well as in a manifestly covariant formulation.
\end{abstract}

\maketitle

\section{Introduction}
\label{sec:intro}

From the approximate chiral symmetry of QCD it is known that the pion
cloud of the nucleon plays a vital role in understanding the nucleon's
long-range structure.
It provides important corrections to static nucleon properties, such as
the mass, magnetic moment and axial charge, and significantly influences
its electric and magnetic charge distributions (for a review see
Ref.~\cite{CBMrev}).
At the quark level, the preferential coupling of a proton to a $\pi^+$
and a neutron provides a natural explanation of the excess of $\bar d$
quarks over $\bar u$ in the proton sea \cite{AWT83}, which has now been
unambiguously established experimentally \cite{NMC,NA51,E866}.
Since it is pseudoscalar, the emission of a pion from a nucleon also
leads to a nontrivial redistribution of the spin and angular momentum
of its quark constituents \cite{Schreiber}, which partially resolves
the proton spin problem \cite{Myhrer}.

More recently, it has been established from the chiral expansion in
QCD that pion cloud contributions to moments of twist-two parton
distributions of the nucleon have a leading nonanalytic (LNA) behavior
characteristic of Goldstone boson loops in chiral perturbation theory
\cite{TMS}.
Since it is determined by the infrared properties of chiral loops,
the LNA behavior is model-independent \cite{Detmold,XDJ,Savage},
and places the physics of the pion cloud on a firm footing in QCD.

In addition to the traditional studies of elastic form factors and
parton distributions, there is a great deal of interest in the more
recently defined generalized parton distributions (GPDs) \cite{GPD}.
As with the ordinary parton distribution functions, the physical
interpretation of GPDs in terms of probability distributions is most
natural on the light-front, or in the infinite momentum frame (IMF)
of time-ordered perturbation theory (TOPT) \cite{DLY}, and one knows
that here too chiral corrections can be very important.
It is timely, therefore, to address the question of how to provide
a consistent derivation of the chiral corrections to all of these
observables on the light-front, and to explain in some detail the
technical differences between these calculations on the light-front
and either in equal-time or covariant formulations.

Historically the realization of chiral symmetry on the light-front has
posed a serious theoretical challenge, and extreme care must usually 
be taken to avoid pathologies associated with so-called zero modes or
spurious end-point singularities in light-front calculations.
Indeed, the LNA behavior of twist-two matrix elements calculated in
the meson cloud model on the light-front (or in the IMF in equal-time
quantization) with a pseudoscalar $\pi N$ interaction \cite{TMS}
appeared to be in conflict with the results from covariant chiral
perturbation theory \cite{XDJ,Savage,XDJ_fy,XDJ_Dfy}, which uses a
pseudovector coupling.
This led to questions being raised \cite{XDJ,Savage} about the
suitability of computing chiral corrections to hadronic matrix
elements in meson cloud models on the light-front
\cite{DLY,MTrho,Speth}.

In other applications, the equivalence of light-front and manifestly
covariant formalisms was demonstrated by Bakker {\em et al.}
\cite{Bakker} for the vector two-point function and pseudoscalar
charge form factor in $1+1$ dimensions.
(For an analogous discussion in QED see Ref.~\cite{MW}.)
Using several different methods, it was shown \cite{Bakker} that
spurious divergences can be avoided when performing loop integrals 
by properly taking into account contributions from the arc used to
close the contour of integration at infinity.
Furthermore, Sawicki \cite{Sawicki} demonstrated for a scalar $\phi^3$
theory that a smooth transition from equal-time perturbation theory
to the light-front can be made without reference to the IMF limit.

In some cases, however, extreme care must be taken when computing
the arc contributions, {\em viz.}, when one encounters so-called
moving poles where the pole in the $k^-$ momentum variable depends
on the $k^+$ integration variable.
In particular, the contributions coming from the end points
$k^+ = 0$ and $k^+ = p^+$ in the $k^+$ range $0 \leq k^+ \leq p^+$
must be taken into account as discussed in Ref.~\cite{Bakker}.
Without these end point contributions the complete equivalence
between the light-front and manifestly covariant results cannot
in general be demonstrated.
It is amusing to see that the corrections we find restore covariance
in the same way as the kinetic mass counter-term does in the work
of Ref.~\cite{MB98}.

In this paper we utilize some of these techniques to demonstrate
the equivalence between equal-time and light-front dynamics for the
interactions of nucleons with pions.
We demonstrate that there is, in fact, no conflict between the results
in equal-time, light-front or covariant frameworks, provided care is
taken when performing loop integrations and if compares consistently
the same theories (with pseudovector or pseudoscalar $\pi N$
interactions).
While a detailed analysis of the twist-two matrix elements in the
different frameworks will be the subject of an upcoming work
\cite{vertex}, here for illustration purposes we consider the
specific example of the self-energy $\Sigma$ of a nucleon dressed
by a pion loop, and examine in particular the model-independent,
nonanalytic behavior of $\Sigma$ in the chiral limit.

In Sec.~\ref{sec:def} we define the Lagrangian for the pseudovector
(PV) $\pi NN$ interaction, and introduce the self-energy for the
dressing of a nucleon by a pion loop.
Although the pseudoscalar (PS) $\pi NN$ interaction does not
preserve chiral symmetry (without the introduction of scalar fields),
for completeness we also consider the pseudoscalar theory in
Appendix~\ref{app:PS}.
We also present a convenient reparametrization of the momentum
dependence in the loop integrations which allows the self-energy
to be expressed entirely in terms of scalar propagators.
The calculation of the self-energy within a covariant framework is
presented in Sec.~\ref{sec:cov} using dimensional regularization.
We derive results for the total $\Sigma$, including finite and
divergent parts, but focus in particular on the structure of the
model-independent, leading (and next-to-leading) nonanalytic
contributions, and recover the standard results of chiral
perturbation theory \cite{ChPT}.

In Sec.~\ref{sec:ET} we examine the self-energy in detail using
time-ordered perturbation theory, both in the familiar rest frame of
the nucleon, and in the IMF, where a probabilistic interpretation is
applicable.
%
%
The computation of $\Sigma$ on the light-front, discussed in
Sec.~\ref{sec:LF}, is closely related to the IMF formulation in
equal-time dynamics.
We verify that in all cases the correct results are obtained
for the nonanalytic contributions.
Finally, in Sec.~\ref{sec:conc} we summarize our findings and
outline future applications of the results.
A presentation of the results for the self-energy with the PS interaction
is given in Appendix~\ref{app:PS}, and the LNA behavior of some relevant
integrals is listed in Appendix~\ref{app:LNA}.

\section{Definitions}
\label{sec:def}

The lowest order $\pi N$ interaction with a pseudovector coupling which
is relevant for the self-energy is defined by the Lagrangian density
\cite{ChPT,Ericson}
\begin{eqnarray}
{\cal L}
&=& {f_{\pi NN} \over m_\pi}
    \left( \bar\psi_N\, \gamma^\mu \gamma_5 \vec\tau\ \psi_N \right)
    \cdot \partial_\mu \vec\phi_\pi\ ,
\label{eq:Lpv}
\end{eqnarray}
where $\psi_N$ and $\vec\phi_\pi$ are the nucleon and pion fields,
$\vec\tau$ is the Pauli matrix operator in nucleon isospin space,
$m_\pi$ is the pion mass, and $f_{\pi NN}$ is the pseudovector
$\pi NN$ coupling constant with $f_{\pi NN}^2/4\pi \approx 0.08$.
The analogous pseudoscalar interaction is given in Appendix~\ref{app:PS}.
Often the PV coupling is expressed in terms of the PS coupling
constant $g_{\pi NN}$,
\begin{eqnarray}
{g_{\pi NN} \over 2M} &=& {f_{\pi NN} \over m_\pi}\ .
\label{eq:cc}
\end{eqnarray}
where $M$ is the nucleon mass, with $g_{\pi NN}^2/4\pi \approx 14.3$.
Using the Goldberger-Treiman relation the $\pi NN$ coupling can also be
expressed in terms of the axial vector charge of the nucleon, $g_A$,
\begin{eqnarray}    
{g_A \over f_\pi} &=& {g_{\pi NN} \over M} ,
\label{eq:GT}
\end{eqnarray}
where $g_A = 1.267$ and $f_\pi \approx 93$~MeV is the pion decay
constant.

The self-energy operator $\widehat\Sigma$ is given by \cite{Hecht}
\begin{eqnarray}
\widehat\Sigma
= i\ \left( {g_{\pi NN} \over 2M} \right)^2
\int{d^4k \over (2\pi)^4}\,
  (\slash{\!\!\!k} \gamma_5 \vec\tau)\,
  {i\, (\slash{\!\!\!p}-\slash{\!\!\!k}+M)
	\over (p-k)^2 - M^2 + i\eps}\,
  (\gamma_5 \slash{\!\!\!k} \vec\tau)\,
{i \over k^2 - m_\pi^2 + i\eps}
\label{eq:SigmaPV}
\end{eqnarray}
and can be decomposed into scalar and vector components according to
\begin{eqnarray}
\widehat\Sigma
= \Sigma_v \slash{\!\!\!p}\ +\ \Sigma_s\ .
\end{eqnarray}
Taking the matrix element of $\widehat\Sigma$ between nucleon states,
the self-energy (of mass shift) of a nucleon with momentum $p$ dressed
by a pion loop with momentum $k$ is given by
\begin{equation}
\Sigma\ =\ {1\over 2} \sum_s \bar u(p,s)\ \widehat\Sigma\ u(p,s)\
	=\ M \Sigma_v\ +\ \Sigma_s\ ,
\label{eq:Sigma}
\end{equation}
where the sum is over the nucleon spins $s$, and we have used the
spinor normalization convention of Bjorken and Drell \cite{BD}.

The expression in Eq.~(\ref{eq:SigmaPV}) can be simplified by
writing the momentum variables in the numerator of the integrand
(\ref{eq:SigmaPV}) in terms of the pion and nucleon propagators
$D_\pi$ and $D_N$, where
\begin{subequations}
\label{eq:D}
\begin{eqnarray}
D_\pi &\equiv& k^2 - m_\pi^2 + i\eps\ ,	\\
D_N   &\equiv& (p-k)^2 - M^2 + i\eps\ .
\end{eqnarray}
\end{subequations}
Rearranging Eqs.~(\ref{eq:D}), and dropping the irrelevant $i \eps$
terms, one can make the replacements
\begin{subequations}
\label{eq:momD}
\begin{eqnarray}
k^2 	  &\to& D_\pi + m_\pi^2\ ,	\\
p \cdot k &\to& {1\over 2} \left( D_\pi - D_N + m_\pi^2 \right)\ .
\end{eqnarray}
\end{subequations}
After applying a trace over the nucleon spins and implementing these
substitutions, the self-energy can be written as
\begin{eqnarray}
\Sigma
&=& -{ 3 i g_{\pi NN}^2 \over 4 M^2 }
\int{d^4k \over (2\pi)^4}
  {1 \over 2M}
  \left[ { 4 M^2 k^2 + 2 k^2\ p\cdot k - 4 (p\cdot k)^2
	   \over D_\pi\ D_N }
  \right]
\label{eq:SigmaPV1}					\\
&=& -{ 3 i g_A^2 \over 4 f_\pi^2 }
\int{d^4k \over (2\pi)^4}
  {1 \over 2M}      
  \left[ 4 M^2 \left( {m_\pi^2 \over D_\pi D_N} + {1 \over D_N} \right)
	+ {2 p \cdot k \over D_\pi}
  \right] ,
\label{eq:SigmaPV2}
\end{eqnarray}
where in (\ref{eq:SigmaPV2}) we have used the Goldberger-Treiman
relation (\ref{eq:GT}).
Note that because the term in Eq.~(\ref{eq:SigmaPV2}) proportional
to $p\cdot k$ is odd in the pion momentum $k$, it will integrate
to zero, provided the ultra-violet regulator does not introduce
additional dependence on $p \cdot k$.
In any case, this should not affect the infrared behavior
of the integrand, and hence not affect the chiral behavior.

\begin{figure}[t]
\includegraphics[height=14cm,angle=270]{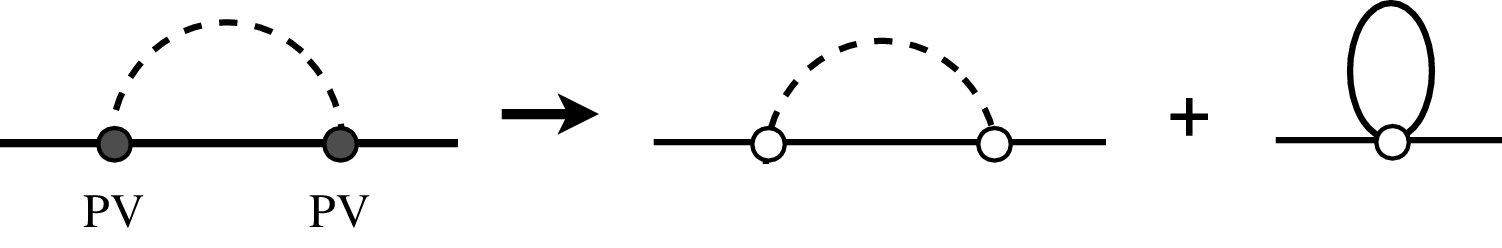}
\caption{Reduction of the self-energy in the pseudovector theory to an
	effective theory of ``scalar nucleons'' and pions (denoted by
	the open blobs at the vertices).}
\label{fig:PV}
\end{figure}

The individual scalar and vector contributions to the self-energy
are given by
\begin{eqnarray}
\Sigma_v
&=& -{ 3 i g_A^2 \over 4 f_\pi^2 }
\int{d^4k \over (2\pi)^4}
  \left[ {1 \over D_N} + {m_\pi^2 \over D_\pi D_N}
	+ {1 \over M^2} {p \cdot k \over D_\pi}
  \right] ,
\label{eq:Sigma_v}					\\
\Sigma_s
&=& -{ 3 i g_A^2 \over 4 f_\pi^2 }
\int{d^4k \over (2\pi)^4}
  M \left[ {1 \over D_N} + {m_\pi^2 \over D_\pi D_N} \right] .
\label{eq:Sigma_s}
\end{eqnarray}
Interestingly, provided that the $p\cdot k$ term in the vector
self-energy does not contribute, the contributions to the total
self-energy from the vector and scalar components are identical,
\begin{eqnarray}
M \Sigma_v = \Sigma_s = {1\over 2}\ \Sigma\ .
\end{eqnarray}
This is not the case, however, for a pseudoscalar theory, as we
discuss in Appendix~\ref{app:PS}.

We can interpret the expression in Eq.~(\ref{eq:SigmaPV2}) as
a reduction of the pseudovector theory to an effective theory
of ``scalar nucleons'' \cite{scalar} and pions, involving
a scalar self-energy and nucleon tadpole diagrams, as in
Fig.~\ref{fig:PV}.
In the following sections we compute the self-energy $\Sigma$
in several ways, including a direct manifestly covariant computation
using dimensional regularization, in equal-time dynamics, both in the
rest frame and in the infinite momentum frame, and on the light-front.

\section{Covariant Formulation with Dimensional Regularization}
\label{sec:cov}

The self-energy can be evaluated in a manifestly covariant manner
using dimensional regularization to regulate the ultra-violet
divergence in the integral.
We can compute the self-energy covariantly by either integrating
the full expression in Eq.~(\ref{eq:SigmaPV1}) or the reduced
result in Eq.~(\ref{eq:SigmaPV2}).
Since the latter is more straightforward, we present results
calculated from the two propagator terms in (\ref{eq:SigmaPV2})
(the $p \cdot k/D_\pi$ term integrates to zero), although we have
verified that identical results are obtained with both expressions.

For the $1/D_\pi D_N$ term, the product of the pion and nucleon
propagators can be reduced using the Feynman parametrization
\begin{subequations}
\label{eq:Feynman}
\begin{eqnarray}
{1\over D_\pi D_N}
&=& {1\over D_\pi - D_N}
    \left( {1\over D_N} - {1\over D_\pi} \right)	\\
&=& \int_0^1 dx { 1 \over \left(x D_\pi + (1-x) D_N\right)^2 } .
\end{eqnarray}
\end{subequations}
Changing variables to $k \to k' = k - (1-x) p$, the denominator in 
the self-energy can be written as
\begin{eqnarray}
x D_\pi + (1-x) D_N
&=& k'^2 + D_{\rm cov} + i\eps\ ,
\end{eqnarray}      
where
\begin{eqnarray}
D_{\rm cov}
&=& -(1-x)^2 M^2 - x m_\pi^2\ .
\end{eqnarray}
Performing a Wick rotation to Euclidean space, the integral over
the pion and nucleon propagators in $d = 4-2\veps$ dimensions,
in the limit $\veps \to 0$, can then be written as
\begin{eqnarray}
\int d^{d}k {1 \over D_\pi D_N}
&=& -i\pi^2
    \left( \gamma + \log\pi - {1 \over \veps}
	 + \int_0^1 dx \log {(1-x)^2 M^2 + x m_\pi^2 \over \mu^2}
	 + {\cal O}(\veps)
    \right) ,
\label{eq:DpiDN}
\end{eqnarray}
where $\mu$ is a mass parameter introduced to give the correct mass
dimensions in $d$ dimensions,
$\Gamma(\veps) = 1/\veps - \gamma + {\cal O}(\veps)$,
and $\gamma \approx 0.577$ is Euler's constant.
For the integral over the tadpole-like $1/D_N$ term, we find
\begin{eqnarray}
\int d^{d}k {1 \over D_N}
&=& -i\pi^2 M^2
    \left( \gamma + \log\pi - {1 \over \veps}
	 + \log{\mu^2 \over M^2}
	 + {\cal O}(\veps)
    \right) ,
\label{eq:Dpi}
\end{eqnarray}
where we have used the recurrence relation for the $\Gamma$ function,
$\Gamma(\veps-1) = \Gamma(\veps)/(\veps-1)$.
The infinitesimal parameter $\veps$ is set to zero at the end of the
calculation, leading to singular results for the integrals, which in
principle can be absorbed into counter-terms when computing observables.
However, our concern here is the finite part of the integrals, and in
particular the LNA behavior of $\Sigma$.

Combining the results in Eqs.~(\ref{eq:DpiDN}) and (\ref{eq:Dpi}),
the self-energy becomes
\begin{eqnarray}
\Sigma_{\rm cov}
&=& -{3 g_A^2 M \over 32 \pi^2 f_\pi^2}
\left\{
  \left( \gamma + \log\pi - {1 \over \veps} + \log{M^2\over\mu^2}
  \right) (M^2 + m_\pi^2)
- M^2 - 2 m_\pi^2
\right.					\nonumber\\
& &
+\ {m_\pi^3 \sqrt{4M^2-m_\pi^2} \over M^2}
  \left( \tan^{-1}{m_\pi \over \sqrt{4M^2-m_\pi^2}}
       + \tan^{-1}{2M^2-m_\pi^2 \over m_\pi\sqrt{4M^2-m_\pi^2}}
  \right)				\nonumber\\
& &
\left.
+\ {m_\pi^4 \over 2 M^2} \log{m_\pi^2 \over M^2}
\right\} .
\label{eq:SigmaCOV}
\end{eqnarray}
It is remarkable that a closed form exists for the complete result
of the self-energy, even in the relativistic formulation
(see also Refs.~\cite{Scherer,Procura}).
Expanding $\Sigma_{\rm cov}$ in powers of $m_\pi/M$ and isolating the
nonanalytic terms (namely, ones which are odd powers or logarithms of
$m_\pi$), the LNA behavior as $m_\pi \to 0$ is given by
\begin{eqnarray}
\Sigma_{\rm cov}^{\rm LNA}
&=& -{3 g_A^2 \over 32 \pi f_\pi^2}
\left( m_\pi^3 + {1 \over 2\pi} {m_\pi^4 \over M} \log m_\pi^2 
	       + {\cal O}(m_\pi^5)
\right) ,
\label{eq:SigmaLNApv}
\end{eqnarray}
where in addition to the ${\cal O}(m_\pi^3)$ term, which agrees with
the established results from chiral perturbation theory \cite{ChPT},
for completeness we have also kept the next order,
${\cal O}(m_\pi^4 \log m_\pi^2)$, nonanalytic term.
This is not the complete contribution to this order, however, as
there exists an ${\cal O}(m_\pi^4 \log m_\pi^2)$ term arising from
diagrams with a pion loop accompanied by a $\Delta$ intermediate
state \cite{Delta}.
In fact, since this contribution depends on $1/(M_\Delta-M)$, where
$M_\Delta$ is the mass of the $\Delta$, rather than on $1/M$ as in
Eq.~(\ref{eq:SigmaLNApv}), it will give the next-to-leading contribution
in the heavy baryon limit.
Note also that because the $1/D_N$ term in Eq.~(\ref{eq:SigmaPV2})
is independent of $m_\pi$, it does not contribute to the nonanalytic
behavior of $\Sigma$, which is solely determined by the $1/D_\pi D_N$
term.

\section{Equal-Time Dynamics}
\label{sec:ET}

While the covariant calculation in the previous section is
straightforward, it is instructive to examine the relative contributions
to the self-energy from the different time orderings of the intermediate
state.
This can be realized using time-ordered perturbation theory, in which
the pion and nucleon propagators are split up into their positive and
negative energy poles.
The relative contributions will naturally depend on the frame of
reference, and we consider two commonly used examples, namely, the
nucleon rest frame and the infinite momentum frame.
Here the 4-dimensional integrals are computed by first performing the
integrations over the energy $k_0$.
The 3-momentum integration over $\bm{k}$ will generally be divergent,
so in order to regularize the integrals we employ a 3-momentum cut-off
on $|\bm{k}|$.
The LNA behavior of the self-energy, which is the primary focus
of this work, will of course be independent of the details of the
ultraviolet regularization.

\subsection{Nucleon Rest Frame}
\label{ssec:RF}

The computation of the self-energy in the nucleon rest frame is most
straightforward in terms of the reduced expression for $\Sigma$ in
Eq.~(\ref{eq:SigmaPV2}).
In contrast to the standard expression in Eq.~(\ref{eq:SigmaPV1})
involving pion momenta in the numerator, the self-energy expressed
solely through pion and nucleon propagators does not receive
contributions from the arc at infinity when performing the $k_0$
integration.
The results are of course identical if one uses the original expression
(\ref{eq:SigmaPV1}), although as we demonstrate below, in that case one
must consider both pole terms and nonzero arc contributions.

Consider first the $1/D_\pi D_N$ term, which when expanded into
positive and negative energy components can be written
\begin{eqnarray}
\int d^4k {1 \over D_\pi D_N}
&=& \int d^3\bm{k} \int_{-\infty}^{\infty} dk_0
{ 1 \over (-2)(\w-i\eps) }
  \left( {1 \over k_0-\w+i\eps} - {1 \over k_0+\w-i\eps} \right)
						\nonumber\\
& & \hspace*{1.2cm}
\times\
{ 1 \over 2(E'-i\eps) }
  \left( {1 \over k_0-E+E'-i\eps} - {1 \over k_0-E-E'+i\eps} \right) ,
\label{eq:DpiDN_RF}
\end{eqnarray}
where in the rest frame the target nucleon has energy $E = M$, the
recoil nucleon energy is $E' = \sqrt{\bm{k}^2 + M^2}$ and the pion
energy is $\w = \sqrt{\bm{k}^2 + m_\pi^2}$.
Multiplying out the terms in the parentheses in Eq.~(\ref{eq:DpiDN_RF}),
the resulting $k_0$ integral has four contributions:
\begin{itemize}

\item	$\Sigma_{\rm ET}^{(+-)}$:
	pion pole in the lower half-plane ($\w-i\eps$) and
	nucleon pole in the upper half-plane ($E-E'+i\eps$);

\item	$\Sigma_{\rm ET}^{(-+)}$:
	pion pole in the upper half-plane ($-\w+i\eps$) and
	nucleon pole in the lower half-plane ($E+E'-i\eps$);

\item	$\Sigma_{\rm ET}^{(++)}$:
	pion pole in the lower half-plane ($\w-i\eps$) and
	nucleon pole in the lower half-plane ($E+E'-i\eps$);

\item	$\Sigma_{\rm ET}^{(--)}$:
	pion pole in the upper half-plane ($-\w+i\eps$) and
	nucleon pole in the upper half-plane ($E-E'+i\eps$),

\end{itemize}
where the superscripts $(\pm\pm)$ refer to the signs of $k_0$
in the pion and nucleon parts of the the energy denominators,
respectively.
Note that the contribution $\Sigma_{\rm ET}^{(+-)}$ corresponds to the
positive energy diagram, while $\Sigma_{\rm ET}^{(-+)}$ is the so-called
``Z-graph''.
Since the terms $\Sigma_{\rm ET}^{(++)}$ and $\Sigma_{\rm ET}^{(--)}$
have both poles in the same half-plane, one can choose the contour of
integration to render their residues zero.
However, in addition to the residues of the poles, the contour
integrations also contain contributions from the arc at infinity,
which must be subtracted,
\begin{equation}
\int_{-\infty}^{\infty}\ =\ \oint_C\ -\ \int_{\rm arc}\ .
\end{equation}
One can verify that closing the contour $C$ in either the upper or
lower half-plane gives the same results for $\Sigma_{\rm ET}^{(+-)}$
and $\Sigma_{\rm ET}^{(-+)}$.
For $\Sigma_{\rm ET}^{(++)}$ the contour can be chosen in the upper
half-plane and for $\Sigma_{\rm ET}^{(--)}$ in the lower half-plane
to exclude the pole contributions.
From the powers of the energy $k_0$ in the numerator and denominators
in Eq.~(\ref{eq:DpiDN_RF}), however, one sees that the arc contributions
will vanish at infinity.
Performing the $k_0$ integration, the four contributions to the
self-energy then become
\begin{subequations}
\label{eq:SigmaETi}
\begin{eqnarray}
\Sigma_{\rm ET}^{(+-)}
&=& -{3 g_A^2 M \over 16 \pi^3 f_\pi^2}
    \int{d^3\bm{k} \over 2E'}\
    {m_\pi^2 \over 2\w}
    \left( {1 \over M-E'-\w} \right) ,		\\
\Sigma_{\rm ET}^{(-+)}
&=& +{3 g_A^2 M \over 16 \pi^3 f_\pi^2}
    \int{d^3\bm{k} \over 2E'}\
    {m_\pi^2 \over 2\w}
    \left( {1 \over M+E'+\w} \right) ,		\\
\Sigma_{\rm ET}^{(++)}
&=& 0\ ,					\\
\Sigma_{\rm ET}^{(--)}
&=& 0\ .
\end{eqnarray}
\end{subequations}

While it does not contribute to the LNA behavior of $\Sigma$,
for completeness we consider also the $1/D_N$ term in
Eq.~(\ref{eq:SigmaPV2}).
Closing the $k_0$ contour integration in either the upper or lower
half-planes, the integral can be written
\begin{eqnarray}
\int d^4k {1 \over D_N}  
&=& \int d^3\bm{k} \int_{-\infty}^{\infty} dk_0
{ 1 \over (k_0-E-E'+i\eps) (k_0-E+E'-i\eps)}	 \nonumber\\
&=& -i \pi \int d^3\bm{k} {1 \over E'}\ .
\end{eqnarray}
Combining all the contributions, the total self-energy in the nucleon
rest frame is
\begin{eqnarray}
\Sigma_{\rm ET}
&=& -{3 g_A^2 M \over 16 \pi^3 f_\pi^2 }
\int{d^3\bm{k} \over 2E'}
{1 \over 2\w}
\left( {4\bm{k}^2 \w + 2E' (\bm{k}^2 + \w^2) \over (E'+\w)^2 - M^2}
\right) .
\end{eqnarray}
The $d^3\bm{k}$ integration can be performed using spherical polar
coordinates, with a high-momentum cut-off $\Lambda$ on $|\bm{k}|$,
yielding the final result for the self-energy,
\begin{eqnarray}
\Sigma_{\rm ET}
&=& -{3 g_A^2 M \over 32 \pi^2 f_\pi^2 }
\Big\{
  2 \Lambda^2 + M^2 + (M^2 + m_\pi^2) \log{M^2 \over 4\Lambda^2}
				\nonumber\\
& & 
+\ {m_\pi^3 \sqrt{4M^2-m_\pi^2} \over M^2}
  \left( \tan^{-1}{m_\pi \over \sqrt{4M^2-m_\pi^2}}
       + \tan^{-1}{2M^2-m_\pi^2 \over m_\pi\sqrt{4M^2-m_\pi^2}}
  \right)			\nonumber\\
& &
+\ {m_\pi^4 \over 2 M^2} \log{m_\pi^2 \over M^2}
\Big\}\, .
\label{eq:SigmaET}
\end{eqnarray}
We observe that the $m_\pi$-dependent terms in Eq.~(\ref{eq:SigmaET})
are identical to those in the covariant calculation of $\Sigma$ in
Eq.~(\ref{eq:SigmaCOV}), with the only differences appearing in terms
that are analytic in $m_\pi$ or which depend on the ultraviolet
regulator.
Not surprisingly, therefore, expanding $\Sigma_{\rm ET}$ in powers
of $m_\pi/M$, the LNA structure in the chiral limit is
\begin{eqnarray}
\Sigma_{\rm ET}^{\rm LNA}
&=& -{3 g_A^2 \over 32 \pi f_\pi^2}
\left( m_\pi^3 + {1 \over 2\pi} {m_\pi^4 \over M} \log m_\pi^2 
	       + {\cal O}(m_\pi^5)
\right) ,
\label{eq:ET_LNA}
\end{eqnarray}
which is consistent with the covariant result in
Eq.~(\ref{eq:SigmaLNApv}).
Note that the ${\cal O}(m_\pi^3)$ term in Eq.~(\ref{eq:ET_LNA})
arises entirely from $\Sigma_{\rm ET}^{(+-)}$, while the
${\cal O}(m_\pi^4 \log m_\pi^2)$ term receives contributions
from both the positive energy and Z-graphs,
\begin{subequations}
\begin{eqnarray}
\Sigma_{\rm ET}^{(+-) \rm LNA}
&=& -{3 g_A^2 \over 32 \pi f_\pi^2}
\left( m_\pi^3\
	+\ {3 \over 4\pi} {m_\pi^4 \over M} \log m_\pi^2\
        +\ {\cal O}(m_\pi^5)
\right) ,					\\
\Sigma_{\rm ET}^{(-+) \rm LNA}
&=& -{3 g_A^2 \over 32 \pi f_\pi^2}
\left( \hspace*{0.8cm}
	-\ {1 \over 4\pi} {m_\pi^4 \over M} \log m_\pi^2\
        +\ {\cal O}(m_\pi^5)
\right) ,
\end{eqnarray}
\end{subequations}
so that combined they reproduce the LNA behavior in
Eq.~(\ref{eq:ET_LNA}).

Had we worked from the original expression for the self-energy in
Eq.~(\ref{eq:SigmaPV1}),
\begin{eqnarray}
\Sigma_{\rm ET}
&=& -{ 3 i g_{\pi NN}^2 \over 4 M^2 } {1 \over (2\pi)^4}
\int d^3\bm{k} \int_{-\infty}^{\infty} dk_0\
{ k_0^3 - k_0 \bm{k}^2 - 2M \bm{k}^2 \over (-2)(\w-i\eps) 2(E'-i\eps) }
						\nonumber\\
& & \hspace*{-1.5cm} \times
\left( {1 \over k_0-\w+i\eps}   - {1 \over k_0+\w-i\eps}   \right)
\left( {1 \over k_0-E+E'-i\eps} - {1 \over k_0-E-E'+i\eps} \right) ,
\end{eqnarray}
then the contributions from the corresponding four cross products
would be
\begin{subequations}
\label{eq:SigmaETj}
\begin{eqnarray}
\Sigma_{\rm ET}^{' (+-)}
&=& -{3 g_A^2 \over 32 \pi^3 f_\pi^2}
\int{d^3\bm{k} \over 2E'} {1 \over 2\w}
\Big\{ { (M-E')^3 + \w^3 - \bm{k}^2 (5M-E'+\w) \over 2(M-E'-\w) }
							\nonumber\\
& & \hspace*{4cm}
-\ { i R_\infty \over \pi } (M-E'+\w)\ +\ {\cal O}(1/R_\infty)
\Big\} ,						\\
\Sigma_{\rm ET}^{' (-+)}
&=& -{3 g_A^2 \over 32 \pi^3 f_\pi^2}
\int{d^3\bm{k} \over 2E'} {1 \over 2\w}
\Big\{ { \w^3 - (M+E')^3 - \bm{k}^2 (-5M-E'+\w) \over 2(M+E'+\w) }
							\nonumber\\
& & \hspace*{4cm}
-\ { i R_\infty \over \pi } (M+E'-\w)\ +\ {\cal O}(1/R_\infty)
\Big\} ,						\\
\Sigma_{\rm ET}^{' (++)}
&=& -{3 g_A^2 \over 32 \pi^3 f_\pi^2}
\int{d^3\bm{k} \over 2E'} {1 \over 2\w}
\Big\{ {1 \over 2} \left[ \w^2 - \bm{k}^2 + \w(M+E') + (M+E')^2 \right]
							\nonumber\\
& & \hspace*{4cm}
+\ { i R_\infty \over \pi } (M+E'+\w)\ +\ {\cal O}(1/R_\infty)
\Big\} ,						\\
\Sigma_{\rm ET}^{' (--)}
&=& -{3 g_A^2 \over 32 \pi^3 f_\pi^2}
\int{d^3\bm{k}\over 2E'} {1 \over 2\w}
\Big\{ {1 \over 2} \left[ \bm{k}^2 - \w^2 + \w(M-E') - (M-E')^2 \right]
							\nonumber\\
& & \hspace*{4cm}
+\ { i R_\infty \over \pi } (M-E'-\w)\ +\ {\cal O}(1/R_\infty)
\Big\} .
\end{eqnarray}
\end{subequations}
Here $R_\infty$ is the magnitude of the energy $k_0$ parametrized
for the arc contribution, $k_0 = R_\infty e^{\pm i\theta}$,
with $\theta$ ranging from 0 to $\pi$, for contours closed in the
upper or lower half-planes, respectively.
In contrast to Eqs.~(\ref{eq:SigmaETi}), the terms
$\Sigma_{\rm ET}^{' (++)}$ and $\Sigma_{\rm ET}^{' (--)}$
are nonzero, with ${\cal O}(R_\infty)$ contributions 
arising from the arc at infinity.
Remarkably, while each individual term in Eqs.~(\ref{eq:SigmaETj})
contains a divergent piece from the arc as $R_\infty \to \infty$,
the total ${\cal O}(R_\infty)$ contribution vanishes once all of the
terms are summed.
One can verify that adding the four terms in Eqs.~(\ref{eq:SigmaETj})
leads to the same result for the total self-energy $\Sigma_{\rm ET}$
as in Eq.~(\ref{eq:SigmaET}).

The nonanalytic behavior of the individual components of the
self-energy in Eqs.~(\ref{eq:SigmaETj}) is given by:
\begin{subequations}
\label{eq:SigmaETjlna}
\begin{eqnarray}
\Sigma_{\rm ET}^{' (+-) \rm LNA}\
=\ -{3 g_A^2 \over 32 \pi f_\pi^2}
  \Big( & m_\pi^3
	& +\ {13 \over 16\pi}\ {m_\pi^4 \over M} \log m_\pi^2
  \Big) ,					\\
\Sigma_{\rm ET}^{' (-+) \rm LNA}\
=\ -{3 g_A^2 \over 32 \pi f_\pi^2}
  \Big( & - {1 \over 2\pi} M m_\pi^2 \log m_\pi^2
	& -\ {5 \over 16\pi}\ {m_\pi^4 \over M} \log m_\pi^2 
  \Big) ,					\\
\Sigma_{\rm ET}^{' (++) \rm LNA}\
=\ -{3 g_A^2 \over 32 \pi f_\pi^2}
  \Big( & {1 \over 2\pi} M m_\pi^2 \log m_\pi^2
	& +\ {1 \over 8\pi}\ {m_\pi^4 \over M} \log m_\pi^2
  \Big) ,					\\
\Sigma_{\rm ET}^{' (--) \rm LNA}\
=\ -{3 g_A^2 \over 32 \pi f_\pi^2}
  \Big( &
	& -\ {1 \over 8\pi}\ {m_\pi^4 \over M} \log m_\pi^2
  \Big) .
\end{eqnarray}
\end{subequations}
Interestingly, while LNA behavior of the positive-energy 
$\Sigma_{\rm ET}^{' (+-)}$ term is ${\cal O}(m_\pi^3)$,
the LNA behavior of the Z-graph $\Sigma_{\rm ET}^{' (-+)}$
actually has a lower order, $\sim m_\pi^2 \log m_\pi$,
which arises partly from the arc contribution.
This cancels, however, with an analogous arc contribution to the
$\Sigma_{\rm ET}^{' (++)}$ component, so that the LNA behavior
of the total is identical to that in Eq.~(\ref{eq:ET_LNA}).
It is clear, therefore, that contributions from the arc at
infinity are vital if the self-energy in the original formulation 
(\ref{eq:SigmaPV1}) is to reproduce the correct behavior of
$\Sigma$ in the chiral limit.
As noted above, the reduced form (\ref{eq:SigmaPV2}) simplifies the
computation of $\Sigma$ considerably by avoiding arc contributions
altogether.

\subsection{Infinite Momentum Frame}
\label{ssec:IMF}

The equal-time calculation allows one to track explicitly the
origins of the various LNA contributions in terms of the respective
time-orderings, which are otherwise obscured in a covariant
calculation.
The price that one pays, however, is that many more time-ordered
diagrams need to be evaluated than in a covariant formulation;
while four graphs for the self-energy is tractable, for other
quantities, such as vertex corrections or multi-loop diagrams,
the number of time orderings quickly escalates.

It was realized some time ago \cite{Weinberg} that by viewing the
system in a Lorentz-boosted frame in which the nucleon is moving
along the $+z$ direction with infinite momentum, many time-ordered
diagrams which contribute in the rest frame are suppressed by powers
of the nucleon momentum, $p_z$, as $p_z \to \infty$.
In particular, diagrams involving backward-moving nucleons
({\em i.e.}, backwards in time) in intermediate states do not
contribute in this limit.
The infinite momentum frame (IMF) therefore provides a simplifying
framework with a much reduced number of diagrams, while at the same
time retaining an intuitive, probabilistic interpretation of particle
production.

The self-energy of the nucleon due to pion loops, as well as vertex and
wave function renormalization, was considered by Drell, Levy and Yan
(DLY) \cite{DLY} using a pseudoscalar $\pi NN$ interaction.
They found, however, that extreme care must be taken to correctly treat
certain cases when the nucleon's momentum fraction carried by the pion
takes its limiting values of 0 or 1, and that naive application of TOPT
rules in the IMF can lead to important contributions to integrals being
omitted.
Here we perform an analogous treatment of the self-energy for the
pseudovector case, and illustrate how so-called ``treacherous'' points
can be avoided.

We begin with the reduced form of the self-energy in
Eq.~(\ref{eq:SigmaPV2}).
For the $1/D_\pi D_N$ term, the integral over the energy $k_0$ can
be decomposed into four contributions as in Eq.~(\ref{eq:DpiDN_RF}).
For the positive energy term, $\Sigma_{\rm IMF}^{(+-)}$, with the
pion pole taken in the lower half-plane and nucleon pole in the upper
half-plane, we parametrize the initial nucleon momentum $p$ and
intermediate nucleon and pion momenta $p'$ and $k$ by \cite{DLY}
\begin{eqnarray}
p &=& (E; \bm{0_\perp}, P)\, , 		\hspace*{2cm}
E  = P + {M^2 \over 2P} + {\cal O}(1/P^2)\, ,
					\nonumber\\
p'&=&(E'; \bm{-k_\perp}, y P)\, ,	\hspace*{1.2cm}
E' = |y| P + {M^2+k_\perp^2 \over 2 |y| P} + {\cal O}(1/P^2)\, ,
					\label{eq:IMFkin} \\
k &=& (\w; \bm{k_\perp}, (1-y) P)\, ,	\hspace*{0.5cm}
\w = |1-y| P + {m_\pi^2+k_\perp^2 \over 2 |1-y| P} + {\cal O}(1/P^2)\, ,
					\nonumber
\end{eqnarray}
where $P \equiv p_z \to \infty$ is the nucleon's longitudinal momentum,
and $y$ is the longitudinal momentum fraction carried by the
intermediate nucleon.
The parametrization (\ref{eq:IMFkin}) respects the momentum
conservation condition $\bm{p} = \bm{p'} + \bm{k}$.
The contribution to the self-energy can then be written
\begin{eqnarray}
\Sigma_{\rm IMF}^{(+-)}
&=& -{3 g_A^2 M \over 16 \pi^3 f_\pi^2}
\int_{-\infty}^\infty dy \int d^2\bm{k_\perp}
{P \over 2E'} {1 \over 2\w} {m_\pi^2 \over (E-E'-\w)}\, ,
\label{eq:IMF+-}
\end{eqnarray}
where we have changed the variable of integration from $k_z$ to $y$.
As we have already shown in the rest frame calculation in
Sec.~\ref{ssec:RF}, the contributions from the arc at infinity to the
$k_0$ integration of the $1/D_\pi D_N$ term are ${\cal O}(1/R_\infty)$
due to the absence of loop momenta in the numerator and the
quadratic dependence on $k_0$ in the denominator.  Only pole terms
contribute, therefore, to the integral over $1/D_\pi D_N$.

For the $y$ integration we need to consider three regions:
$y < 0$, $0 < y < 1$ and $y > 1$.
For $y < 0$, which corresponds to the intermediate nucleon moving
backward (in space), from the momenta parametrizations in
Eq.~(\ref{eq:IMFkin}) the energy denominator in
Eq.~(\ref{eq:IMF+-}) is
\begin{equation}
E-E'-\w = 2 y P + {\cal O}(1/P) \hspace*{4cm} (y < 0) .
\end{equation}
Counting powers of the large momentum $P$, one sees that the integral
vanishes as $P \to \infty$.
In the region $0 < y < 1$ the energy denominator becomes
\begin{equation}
\hspace*{0.7cm}
E-E'-\w = - {k_\perp^2 + M^2 (1-y)^2 + m_\pi^2 y \over 2y(1-y) P}
\hspace*{2.3cm} (0 < y < 1) ,
\end{equation}
which yields a nonzero contribution to (\ref{eq:IMF+-})
in the limit $P \to \infty$.
Finally, when $y > 1$, which corresponds to the pion moving backward,
one has
\begin{equation}
E-E'-\w = 2 (1-y) P + {\cal O}(1/P) \hspace*{3cm} (y > 1) ,
\end{equation}
which again gives a suppressed contribution as $P \to \infty$.
Thus the only nonzero contribution is from the diagram where both
the nucleon and pion are forward moving,
{\em i.e.}, $y > 0$ and $1-y > 0$, in which case the self-energy
for the positive energy diagram is
\begin{eqnarray}
\Sigma_{\rm IMF}^{(+-)}
&=& {3 g_A^2 M \over 32 \pi^2 f_\pi^2}
\int_0^1 dy \int_0^{\Lambda_\perp^2} dk_\perp^2\
{ m_\pi^2 \over k_\perp^2 + M^2 (1-y)^2 + m_\pi^2 y }\, ,
\label{eq:SigmaIMF+-}
\end{eqnarray}
where an ultraviolet cut-off $\Lambda_\perp$ is introduced to
regulate the $k_\perp$ integration.

For the Z-graph, the intermediate nucleon and pion momenta are
parametrized according to \cite{DLY}
\begin{equation}
p' = (E'; \bm{k_\perp}, -y P)\, , \hspace*{1.2cm}
k  = (\w; -\bm{k_\perp}, -(1-y) P)\, ,
\label{eq:IMFZkin}
\end{equation}
satisfying the momentum conservation condition
$\bm{p} + \bm{p'} + \bm{k} = 0$.
The contribution to the self-energy is then given by
\begin{eqnarray}
\Sigma_{\rm IMF}^{(-+)}
&=& {3 g_A^2 M \over 16 \pi^3 f_\pi^2}
\int_{-\infty}^\infty dy \int d^2\bm{k_\perp}
{P \over 2E'} {1 \over 2\w} {m_\pi^2 \over (E+E'+\w)}\, ,
\label{eq:IMF-+}
\end{eqnarray}
where again arc contributions at infinity are suppressed.
Unlike for the $\Sigma_{\rm IMF}^{(+-)}$ case above, however,
since each of the ${\cal O}(P)$ terms in the energy denominator
add, the sum $E+E'+\w = {\cal O}(P)$ for all $y$.
Counting the large momentum factors in Eq.~(\ref{eq:IMF-+})
reveals that the Z-graph contribution to the self-energy is
\begin{eqnarray}
\Sigma_{\rm IMF}^{(-+)}
&=& {\cal O}(1/P^2)\, ,
\end{eqnarray}
and thus vanishes in the limit $P \to \infty$.
Moreover, since there are no arc contributions for the $1/D_\pi D_N$
term in the equal-time dynamics, the terms with both poles in the
upper or lower half-plane will be zero,
$\Sigma_{\rm IMF}^{(++)} = \Sigma_{\rm IMF}^{(--)} = 0$,
as for the rest frame calculation in Eq.~(\ref{eq:SigmaETi}).

For the $1/D_N$ nucleon tadpole term in Eq.~(\ref{eq:SigmaPV2}),
the integration can be performed in an analogous way, using the
parametrization of the momenta as in Eq.~(\ref{eq:IMFkin}).
Although it does not contain nonanalytic structure in $m_\pi$,
it is still instructive to examine its computation in the IMF,
and to demonstrate the equivalence of the results with those
of the formalisms.
Using the fact that the integrand is symmetric in $y \to -y$,
the integral of $1/D_N$ can be written
\begin{eqnarray}
\int d^4k {1 \over D_N}  
&=& -2\pi^2 i
\int_0^\infty dy \int_0^{\Lambda_\perp^2} dk_\perp^2
{ 1 \over \sqrt{y^2 + (M^2+k_\perp^2)/P^2} } ,
\end{eqnarray}
where $\Lambda_\perp$ cuts off the large transverse momenta in the
$k_\perp$ integration.
After integrating over $k_\perp$, we note that because of the simpler
structure of the denominator, we need only consider a single region
of $y$ integration.
Since the $y$ integration is also divergent, we introduce a
large-$y$ cut-off parameter, $\lambda$, 
\begin{eqnarray}
\int d^4k {1 \over D_N}
&=& -2\pi^2 i
\left( {\Lambda_\perp^2 \over 2}
      + \Lambda_\perp^2 \log{2\lambda P \over \Lambda_\perp}
      + M^2 \log{M \over \Lambda_\perp}
\right)\, ,
\end{eqnarray}
with $\lambda \to \infty$.
One could also perform the $y$ integration first, as in DLY \cite{DLY}
for the pseudoscalar model, with yields the same results.

Combining the $\pi N$ loop diagram with the nucleon tadpole, the total
self-energy in the IMF can be written
\begin{eqnarray}
\Sigma_{\rm IMF}
&=& -{3 g_A^2 M \over 32 \pi^2 f_\pi^2 }
\Big\{
  \Lambda_\perp^2
  \left( 1 + \log{4 \lambda^2 P^2 \over \Lambda_\perp^2} \right)
  - 2 m_\pi^2 + (M^2+m_\pi^2) \log{M^2 \over \Lambda_\perp^2}
				\nonumber\\
& & 
+\ {m_\pi^3 \sqrt{4M^2-m_\pi^2} \over M^2}
  \left( \tan^{-1}{m_\pi \over \sqrt{4M^2-m_\pi^2}}
       + \tan^{-1}{2M^2-m_\pi^2 \over m_\pi\sqrt{4M^2-m_\pi^2}}
  \right)			\nonumber\\
& &
+\ {m_\pi^4 \over 2 M^2} \log{m_\pi^2 \over M^2}
\Big\}\, ,
\label{eq:SigmaIMF}
\end{eqnarray}
which is identical to the rest frame expression (\ref{eq:SigmaET})
in its nonanalytic structure.
Expanding in $m_\pi$ then gives the LNA behavior of the self-energy
in the IMF, which arises solely from the positive energy $\pi N$
loop contribution,
\begin{eqnarray}
\Sigma_{\rm IMF}^{\rm LNA}\ =\ \Sigma_{\rm IMF}^{(+-) \rm LNA}
&=& -{3 g_A^2 \over 32 \pi f_\pi^2}
\left( m_\pi^3 + {1 \over 2\pi} {m_\pi^4 \over M} \log m_\pi^2
               + {\cal O}(m_\pi^5)
\right)\, .
\end{eqnarray}
The results differ, not surprisingly, from those of DLY \cite{DLY}
for the pseudoscalar $\pi N$ interaction.
In that case the calculation is performed using the original
expression for $\Sigma$ in Eq.~(\ref{eq:SigmaPS1}) below, where the
presence of loop momenta in the numerator means that the main
contribution to the self-energy arises from the infinitesimal
end-point regions $-\eps < y < \eps$ and $1-\eps < y < 1+\eps$
\cite{DLY} (see Appendix~\ref{app:PS}).
For the more consistent pseudovector theory, working with the
reduced form (\ref{eq:SigmaPV2}) allows one to avoid this problem.

\section{Light-Front Dynamics}
\label{sec:LF}

A formalism which is similar to TOPT in the IMF is light-front, where
the fields are quantized along the light-cone, $(ct)^2 - z^2 = 0$.
The advantage of this formulation is that, as with the IMF in
equal-time, many diagrams are suppressed, but the suppression is
realized in any reference frame.
However, as also in the IMF calculation, care needs to be taken to
correctly include contributions from end-point regions corresponding
to pion light-front momentum fractions equal to 0 or 1.

We define four-momenta on the light-front as
$v = (v^+, v^-, \bm{v_\perp})$, with the plus/minus components
$v^\pm = v_0 \pm v_z$.
Note that particles on the light-front are on their mass shells,
while the minus components of momenta are not conserved at vertices.
In terms of light-front variables, the self-energy in
Eq.~(\ref{eq:SigmaPV2}) can be written as
\begin{eqnarray}
\Sigma_{\rm LF}
&=& -{ 3 i g_A^2 M \over 4 f_\pi^2 }
{1 \over (2\pi)^4}
\int dk^+ dk^- d^2\bm{k_\perp}
  \left( {m_\pi^2 \over D_\pi D_N} + {1 \over D_N} \right) ,
\label{eq:SigmaLFdef}
\end{eqnarray}
where we have dropped the term odd in $k$.

Choosing the $\bm{p_\perp}=0$ frame, and performing the $k^-$
integration, the integral of the $1/D_\pi D_N$ term can be written
\begin{eqnarray}
\int dk^+ dk^- d^2\bm{k_\perp} {1 \over D_\pi D_N}
&=& {1 \over p^+}
\int_{-\infty}^{\infty} {dx \over x (x-1)} d^2\bm{k_\perp}
							\nonumber\\
& & \times
\int dk^-
\left( k^- - {k_\perp^2+m_\pi^2 \over xp^+} + {i\eps \over xp^+}
\right)^{-1}						\nonumber\\
& & \times
\left( k^- - {M^2 \over p^+} - {k_\perp^2+M^2 \over (x-1)p^+}
	   + {i\eps \over (x-1)p^+}
\right)^{-1}						\nonumber\\
&=& 2\pi^2 i \int_0^1 dx\ dk_\perp^2\
{1 \over k_\perp^2 + (1-x) m_\pi^2 + x^2 M^2} ,
\label{eq:SigmaLF_DpiDN}
\end{eqnarray}
where $x = k^+/p^+$ is the plus momentum fraction of the nucleon
carried by the pion.
There is no arc contribution here due to the quadratic $k^-$
dependence in the denominator.
Thus, for the region $x < 0$ or $x > 1$, where both poles of $k^-$
are located either in the lower or upper half plane, respectively,
the integral of $1/D_\pi D_N$ vanishes.
Consequently, only the region $0 < x < 1$ contributes to the integration.
Note that the result in (\ref{eq:SigmaLF_DpiDN}) identical to the IMF
result for $\Sigma_{\rm IMF}^{(+-)}$ in Eq.~(\ref{eq:SigmaIMF+-}),
provided the $k_\perp$ integration is regulated by the same
high-momentum cut-off $\Lambda_\perp$ as in (\ref{eq:SigmaIMF+-}).

For the $1/D_N$ term care needs to be taken in computing the arc
contribution because of the dependence on $k^+$ in the pole of $k^-$.
Namely, the $k^-$ pole is moving as $k^+$ changes, which leads to a
treacherous point as has been discussed previously in the literature
\cite{Bakker,MW,MB98,LB,PB,Yan}.
To see this point we can change the integration variable $k-p \to k$
and rewrite the $1/D_N$ term as
\begin{eqnarray}
\int d^4k {1 \over D_N}
&=& \int d^4k {1 \over k^2 - M^2 + i\eps}	\nonumber\\
&=& {1\over 2} \int d^2\bm{k_\perp} \int {dk^+ \over k^+}
\int dk^- \left( k^- - {k_\perp^2+M^2 \over k^+} + {i\eps \over k^+}
\right)^{-1}\ .
\label{eq:DN_LF}
\end{eqnarray}
Note here that the $k^-$ pole
({\em i.e.}, $k^- = (k_\perp^2+M^2)/k^+ - i\eps/k^+)$
depends on $k^+$.
Not only does the position of the pole depend on the sign of $k^+$,
but also the pole moves to infinity as $k^+ \to 0$, changing the
degree of divergence of the $k^-$ integral from logarithmic to linear.
At the point $k^+ = 0$ the $k^-$ integral is linearly divergent, not
logarithmically divergent as one would naively expect from
Eq.~(\ref{eq:DN_LF}) for the $k^+ > 0$ and $k^+ < 0$ regions.
The computation of the $1/D_N$ term is thus highly nontrivial in
light-front dynamics.

The details of this treacherous point have been discussed by
Bakker {\em et al.} \cite{Bakker}.
Following Ref.~\cite{Bakker}, we use the light-front cylindrical
coordinates ($k^+ = r \cos\phi$, $k^- = r \sin\phi$) to perform the
$k^+$ and $k^-$ integration as follows:
\begin{eqnarray}
\int dk^+ dk^-
  {1 \over k^+ k^- - k_\perp^2 - M^2 + i\eps}
&=& \int_0^\infty dr\ r \int_0^{2\pi} d\phi
  \left( {r^2 \sin\phi \cos\phi - k_\perp^2 - M^2 + i\eps} \right)^{-1}
\nonumber\\
&=& -4\pi
\left[
  \int_0^{r_0} dr {r \over \sqrt{r_0^4-r^4}}
+ i \lim_{R\to\infty} \int_{r_0}^R dr {r \over \sqrt{r^4-r_0^4}}
\right]							\nonumber\\
&=& \lim_{R\to\infty}
\left( -\pi^2 + 2\pi i\ \log{r_0^2 \over R^2} + {\cal O}(1/R^4)
\right) ,
\end{eqnarray}
where $r_0 = \sqrt{2 (k_\perp^2 + M^2)}$.
The result contains the same $\log(k_\perp^2+M^2)$ term in the
integrand as in the IMF calculation above.
The $k_\perp$ integration is then straightforward, and adding the
two terms in Eq.~(\ref{eq:SigmaLFdef}) gives the total light-front
self-energy
\begin{eqnarray}
\Sigma_{\rm LF}
&=& -{3 g_A^2 M \over 32 \pi^2 f_\pi^2 }
\left\{
  \Lambda_\perp^2
+ m_\pi^2 \log{M^2 \over \Lambda_\perp^2}
- M^2 \log\left(1+{\Lambda_\perp^2 \over M^2}\right)
- \Lambda_\perp^2 
  \log\left( {\Lambda_\perp^2 + M^2 \over R^2 e^{-i\pi/2}} \right)
\right.				\nonumber\\
& & 
+\ {m_\pi^3 \sqrt{4M^2-m_\pi^2} \over M^2}
  \left( \tan^{-1}{m_\pi \over \sqrt{4M^2-m_\pi^2}}
       + \tan^{-1}{2M^2-m_\pi^2 \over m_\pi\sqrt{4M^2-m_\pi^2}}
  \right)			\nonumber\\
& &
\left.
+\ {m_\pi^4 \over 2 M^2} \log{m_\pi^2 \over M^2}
\right\}\, .
\label{eq:SigmaLF}
\end{eqnarray}
The nonanalytic part of $\Sigma_{\rm LF}$ is identical to the results
of the covariant and equal-time (rest frame and IMF) calculations,
\begin{eqnarray}
\Sigma_{\rm LF}^{\rm LNA}
&=& -{3 g_A^2 \over 32 \pi f_\pi^2}
\left( m_\pi^3 + {1 \over 2\pi} {m_\pi^4 \over M} \log m_\pi^2 
	       + {\cal O}(m_\pi^5)
\right) ,
\end{eqnarray}
demonstrating the equivalence of the light-front formalisms to the
equal time and covariant formulations.

\section{Conclusions}
\label{sec:conc}

We have demonstrated the equivalence of the covariant, equal-time
(both in the rest frame and the IMF), and light-front formalisms
in computing the renormalization of the bare nucleon by pion loops.  
As a specific example, we have focussed on the self-energy of the 
nucleon, which finds applications in computations of pion cloud
corrections to deep inelastic structure functions, form factors,
and other observables.
While this equivalence is expected, it has not to our knowledge been
demonstrated explicitly in the literature.
The calculations involve some non-trivial aspects, especially in the
IMF and LF formulations, which require special care when dealing with
end-point singularities, corresponding to momentum fractions approaching
0 or 1.

We focused on the chirally preferred case of pseudovector coupling,
although the comparison with pseudoscalar coupling (discussed in
Appendix~\ref{app:PS}), which was employed in the classical work of
Drell, Levy and Yan (DLY), reveals several interesting features.  
The leading nonanalytic behavior of the self-energy in the latter
case is incorrect, with a term behaving as $m_\pi^2 \ln m_\pi$. 
In addition, whereas in the pseudovector case the scalar and vector
pieces are equal and have the same LNA behavior, in the pseudoscalar 
theory they have even lower order (incorrect) nonanalytic terms of 
order $m_\pi$, which cancel in the full combination
$M \Sigma_v + \Sigma_s$.
Finally, we note that the technical problems encountered by DLY, with
critical contributions from the regions $-\epsilon < x < \epsilon$
and $1-\epsilon < x < 1+\epsilon$, are peculiar to the pseudoscalar
case in the light-front formalism.

There are a number of applications of the methodology developed here
which will be important to pursue in future.
As well as the LNA behavior of the vertex renormalization, $Z_1$,
it is important to compute the LNA behavior of the moments of the
twist-two parton distribution functions.
In particular, one will be able to explore within the LF formalism
the results obtained by DLY \cite{DLY} in the IMF, as well as the
application to this problem of effective field theory
\cite{XDJ_Dfy,Savage}.
This should permit a satisfactory resolution of the discrepancy between
the results of IMF and rest frame equal-time calculations cited by 
Chen and Ji~\cite{XDJ_Dfy}.
It will also pave the way for a consistent interpretation of the
physics of the pion cloud at the partonic level and for the development
of realistic chiral models of hadron structure on the LF suitable for
discussing form factors, parton distribution functions and GPDs.

\section*{Acknowledgements}

This work was supported by the DOE contract No. DE-AC05-06OR23177,
under which Jefferson Science Associates, LLC operates Jefferson Lab.
We acknowledge partial support from the Triangle Nuclear Theory fund
and the NCSU theory group fund (DOE contract No. DE-FG02-03ER41260)

\newpage
\appendix
\section{Self-energy for the Pseudoscalar Theory}
\label{app:PS}

Although the pseudoscalar Lagrangian is not invariant under chiral
transformations without the introduction of a scalar field (as in the
linear sigma model, for example \cite{TW}), since this theory is often
discussed in the literature, for completeness we discuss here the
results for the PS theory.
This also allows us to contrast a number of important features
pertinent to the pseudovector and pseudoscalar calculations.

The lowest order Lagrangian density for a PS $\pi N$ interaction
relevant for the self-energy is given by
\begin{eqnarray}
{\cal L}^{\rm PS}
&=& - g_{\pi NN}
    \left( \bar\psi_N\, i \gamma_5 \vec\tau\, \psi_N \right)
    \cdot \vec\phi_\pi\ ,
\label{eq:Lps}
\end{eqnarray}
where $g^2_{\pi NN}/4\pi \approx 14.3$ is the PS coupling constant.
For on-shell nucleons obeying the free Dirac equation the PS and PV
Lagrangians (see Eq.~(\ref{eq:Lpv})) give identical results for matrix
elements, provided the couplings are related by Eq.~(\ref{eq:cc}).
For bound or off-shell nucleons, the PS and PV interactions lead to
different results.
However, the couplings $f_{\pi NN}$ and $g_{\pi NN}$ are still related
through Eq.~(\ref{eq:cc}) since these are defined at the nucleon (and
pion) poles.

\begin{figure}
\includegraphics[height=14cm,angle=270]{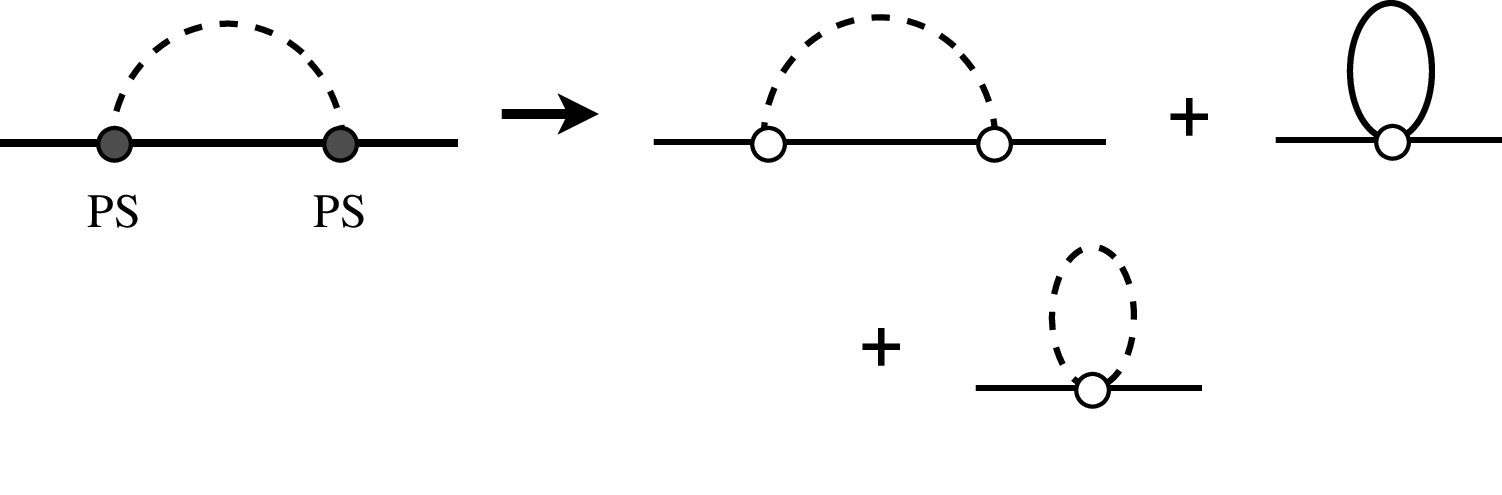}
\caption{Reduction of the self-energy in the pseudoscalar theory to an
	effective theory of ``scalar nucleons'' and pions (denoted by
	the open blobs at the vertices).}
\label{fig:PS}
\end{figure}

The self-energy operator for the PS coupling is given by
\begin{equation}
\widehat\Sigma^{\rm PS}
= i\ g_{\pi NN}^2
\int{d^4k\over (2\pi)^4}
  (\gamma_5 \vec\tau)
  {i\, (\slash{\!\!\!p}-\slash{\!\!\!k}+M)
	\over (p-k)^2 - M^2 + i\eps}
  (\gamma_5 \vec\tau)
{i \over k^2 - m_\pi^2 + i\eps}\ .
\label{eq:SigmaPS}
\end{equation}
Taking the spin trace and using Eq.~(\ref{eq:D}) to replace
the momentum dependence in the numerator by the nucleon and
pion propagators gives for the self-energy
\begin{eqnarray}
\Sigma^{\rm PS}
&=& -3 i g_{\pi NN}^2
\int{d^4k \over (2\pi)^4}
  {1 \over 2M}
  {2 p\cdot k \over D_\pi D_N}
\label{eq:SigmaPS1}			\\
&=& -{3 i g_A^2 M \over 2 f_\pi^2}
\int{d^4k \over (2\pi)^4}
  \left[ {m_\pi^2 \over D_\pi D_N}
	+ {1 \over D_N}
	- {1 \over D_\pi}
  \right]\ .
\label{eq:SigmaPS2}
\end{eqnarray}
The transformed effective PS theory is represented in Fig.~\ref{fig:PS},
where the self-energy now contains contributions from a scalar nucleon
self-energy and a nucleon tadpole diagram, as in the PV theory, as well
as a pion tadpole term.

The individual vector and scalar contributions to $\Sigma$ are
given by
%
\begin{eqnarray}
\Sigma^{\rm PS}_v
&=& -3i g_{\pi NN}^2     
\int{d^4k\over (2\pi)^4}
{1 \over 2M^2}
\left[ {m_\pi^2-2M^2 \over D_\pi D_N} - {1 \over D_\pi} + {1 \over D_N}
\right]\ ,		\\
\Sigma^{\rm PS}_s
&=& -3i g_{\pi NN}^2
\int{d^4k\over (2\pi)^4}
{M \over D_\pi D_N}\ ,
\end{eqnarray}
%
which when combined recover the full expression in
Eq.~(\ref{eq:SigmaPS2}).

The result of the covariant calculation of the full self-energy
$\Sigma^{\rm PS}$ using dimensional regularization is
\begin{eqnarray}
\Sigma^{\rm PS}_{\rm cov}
&=& -{3 g_A^2 M \over 32 \pi^2 f_\pi^2}
\left\{
  \left( \gamma + \log\pi - {1 \over \veps} + \log{M^2\over\mu^2}
  \right) M^2
- (M^2 + m_\pi^2)
\right.                                 \nonumber\\
& &
+\ {m_\pi^3 \sqrt{4M^2-m_\pi^2} \over M^2}
  \left( \tan^{-1}{m_\pi \over \sqrt{4M^2-m_\pi^2}}
       + \tan^{-1}{2M^2-m_\pi^2 \over m_\pi\sqrt{4M^2-m_\pi^2}}
  \right)                               \nonumber\\
& &
\left.
-\ m_\pi^2 \log{m_\pi^2 \over M^2}   
+\ {m_\pi^4 \over 2 M^2} \log{m_\pi^2 \over M^2}   
\right\} .
\label{eq:SigmaPS_COV}
\end{eqnarray}
Expanding in powers of $m_\pi$, the leading nonanalytic behavior
of the self-energy in the PS theory is given by
\begin{eqnarray}
\Sigma^{\rm PS}_{\rm LNA}
&=& {3 g_A^2 \over 32 \pi f_\pi^2}
\left( {M \over \pi} m_\pi^2 \log m_\pi^2
      - m_\pi^3
      - {m_\pi^4 \over 2 \pi M^2} \log{m_\pi^2 \over M^2}
      + {\cal O}(m_\pi^5)
\right)\ .
\label{eq:SigmaLNAps}
\end{eqnarray}
Here the ${\cal O}(m_\pi^3)$ term is identical to the leading term
in the PV self-energy in Eq.~(\ref{eq:SigmaLNApv}).
This is not surprising since the $m_\pi^2/D_\pi D_N$ term in the
integrand, which produces the $m_\pi^3$ behavior, is common to
both the PV and PS self-energies.
However, the presence of the pion tadpole in $\Sigma^{\rm PS}$
(arising from the $1/D_\pi$ term in the integrand) leads to an
additional contribution proportional to $m_\pi^2 \log m_\pi^2$,
which is of lower order than the $m_\pi^3$ term.

Moreover, if we consider the leading nonanalytic behavior of the
vector and scalar parts separately,
\begin{eqnarray}
\Sigma^{\rm PS}_{v, \rm LNA}
&=& {3 g_A^2 \over 32 \pi f_\pi^2} {1 \over M}
\left[ 2M^2 m_\pi
     + {2M \over \pi} m_\pi^2 \log m_\pi^2
     - m_\pi^3
     + \cdots
\right]\ ,			\\
\Sigma^{\rm PS}_{s, \rm LNA}
&=& -{3 g_A^2 \over 32 \pi f_\pi^2}
\left[ 2M^2 m_\pi
     + {M \over \pi} m_\pi^2 \log m_\pi^2
     + \cdots
\right]\ ,
\end{eqnarray}
we find that, in contrast to the PV case, the leading order terms in
both $\Sigma^{\rm PS}_{v, \rm LNA}$ and $\Sigma^{\rm PS}_{s, \rm LNA}$
are ${\cal O}(m_\pi)$!
While these terms cancel in the total $\Sigma^{\rm PS}_{\rm LNA}$,
this demonstrates that the LNA behavior of $\Sigma^{\rm PS}$ is very
different to that for the PV theory.

Using the original formulation as in Eq.~(\ref{eq:SigmaPS1}),
Drell, Levy and Yan \cite{DLY} computed the mass shift in the
IMF formulation, finding that the entire contribution arises
from the infinitesimal end-point regions $-\eps < y < \eps$ and
$1-\eps < y < 1+\eps$ for the positive energy and $Z$-graph
contributions, respectively.
Moreover, the two terms separately diverge as $P \to \infty$
or $\eps \to 0$, but the sum of the two becomes independent
of $P$ and $\eps$.
While the results are ultimately the same, this illustrates the
fact that the original formulation (\ref{eq:SigmaPS1}) leads to a
somewhat more tortuous path, at least in the IMF (and light-front)
treatments than with the reduced form (\ref{eq:SigmaPS2}).

\section{LNA Behavior of Integrals}
\label{app:LNA}

In this appendix we summarize the leading nonanalytic behavior of
various integrals which enter into calculations of the self-energy
in the PV and PS theories.
The relevant integrands are ones which involve a product of pion
and nucleon propagators, one pion propagator (for the pion tadpole
diagram in Fig.~\ref{fig:PV}), or one nucleon propagator (for the
nucleon tadpole diagrams in Figs.~\ref{fig:PV} and \ref{fig:PS}).

For the product of pion and nucleon propagators, one finds leading
terms of order $m_\pi$ and $m_\pi^2 \log m_\pi^2$,
\begin{eqnarray}
\left. \int d^4k\ {m_\pi^2 \over D_\pi D_N} \right|_{\rm LNA}
&=& -i \pi^3 {m_\pi^3 \over M}
    \left( 1 + {1 \over 2\pi} {m_\pi \over M} \log m_\pi^2 \right)
+ {\cal O}(m_\pi^5)\ .
\end{eqnarray}
The pion tadpole diagram involves the integral over a single pion
propagator, and contributes to the nonanalytic behavior only at a
higher order,
\begin{eqnarray}
\left. \int d^4k\ {1 \over D_\pi} \right|_{\rm LNA}
&=& -i \pi^2 m_\pi^2 \log m_\pi^2\ .
\end{eqnarray}
Note that there are no higher order contributions in $m_\pi$
for this term.
Finally, since it is independent of the pion mass, integrals
involving nucleon propagators only will give zero nonanalytic
contributions,
\begin{eqnarray}
\left. \int d^4k\ {1 \over D_N^n} \right|_{\rm LNA}
&=& 0\ ,
\end{eqnarray}
for any power $n$.

\newpage


\begin{thebibliography}{99}
\bibitem{scalar}
By ``scalar nucleon'' we mean that there is no momentum in the numerator
of the self-energy $\Sigma$, as in a theory of pions and spin-0 nucleons.
\end{thebibliography}
\end{document}